\documentstyle[12pt]{article}

\newcommand{\LaSrCuO}  {{LaSr$_{\rm x}$Cu$_{2-{\rm x}}$O$_4$\ }}
\newcommand{\YBCO}[1]  {{YBa$_2$Cu$_3$O$_{#1}$\ }}

\newcommand{\cd}[1] {{c^\dagger_{#1}}}
\newcommand{\cu}[1] {{c^{\phantom{\dagger}}_{#1}}}
\newcommand{\pk}[1] {{p^\dagger_{#1}}}
\newcommand{\pu}[1] {{p^{\phantom{\dagger}}_{#1}}}
\newcommand{\dd}[1] {{d^\dagger_{#1}}}
\newcommand{\du}[1] {{d^{\phantom{\dagger}}_{#1}}}
\newcommand{\sd}[1] {{s^\dagger_{#1}}}
\newcommand{\su}[1] {{s^{\phantom{\dagger}}_{#1}}}

\newcommand{\Figure}[1]
{
	\includegraphics{#1}
	\vspace{9cm}
}

\newcommand{\SmallProceedingsFigure}[1]
{
	\includegraphics{#1}
	\vspace{7.0cm}
}

\begin{document}

\bibliographystyle{PhysRev}

\title{
Aspects of the Spin Dynamics in the Cuprate Superconductors
}

\author{
H. Monien\\
Institute for Theoretical Physics\\
University of California\\
Santa Barbara, CA 93106, USA
}

\date{\today}

\vspace{0.2cm}
\maketitle
\begin{center}
 {\bf\large Abstract:}
\end{center}
{
One of the interesting aspects of the CuO superconductors is that
superconductivity is happening so close to the antiferromagnetic state.
The nuclear magnetic resonance and the recent neutron scattering experiments
clearly indicate that magnetic correlations persist in to the heavily
doped regime. In this paper we will discuss some of the details of the
coupling of the nuclear magnetic spin to the conduction electron spins.
Furthermore we will show that a simple band structure can explain the
recent neutron scattering data in the \LaSrCuO material for the optimal
concentration of $x\approx 0.15$ if the lifetime effects are included.
\/}

\newpage

\section{Introduction}

 From the theoretical perspective one of the most interesting aspects of the
CuO high temperature superconductors is that they belong to a class of
materials which show strong electronic correlations. In the undoped materials
these correlations lead to the antiferromagnetic ordering of the electronic
spins. This is why Anderson \cite{Anderson} pointed out early on that the
physics of the CuO superconductors is best described in terms of the Hubbard
model.
The model Hamiltonian contains the strong on site Coulomb repulsion
and a hopping term describing the kinetic energy. The theoretically
fascinating aspect is how to obtain superconductivity from a purely
repulsive interaction.
Very close to half filling where this model describes the insulating
Heisenberg antiferromagnet it is extremely well understood and the neutron
scattering experiments \cite{Kastner} and theoretical calculations, e.g.
\cite{Chakravarty}, are in beautyful agreement. Away from half
filling the situation is much less clear.
Except in one dimensions and recently infinite dimensions the physics
of the Hubbard model is not understood to a degree that one knows the
correlation  functions quantitatively. Even today small cluster calculations
and approximations which are uncontrolled in the relevant regime
are the only way one is able to achieve some
progress in the strongly correlated regime away from half filling.

Under these circumstances it is useful to know what limitations one can
obtain from experiments for a theory for the CuO superconductors. The
constraints obtained from experiments might tell us if a theoretical
model like the Hubbard model contains even qualitatively the right
physics to describe the low energy excitations of the ``real materials''
or if some additional physics is required to understand the basic properties
of the high temperature superconductors. It is quite clear that these
materials are close to many instabilities which might all be important.
Thus, choosing to concentrate on a particular degree of freedom of the
system, one might completely miss the relevant physics. A ``worst case
scenario'' from a theorists point of view would be that all the possible
degrees of freedom of the system contribute to the most interesting of
all the instabilities of the system: the superconducting state.

In order
to see ``a way through the jungle'', I will concentrate on the magnetic
properties of the copper oxide superconductors which are very interesting
in themselves. The magnetic properties of the CuO materials were studied in
great detail by the nuclear magnetic resonance techniques by many groups
and by neutron scattering experiments which are very difficult to perform
in these compounds in comparison with standard materials. In both types
of experiments one would like to obtain information about the dynamical
structure factor of the electronic spins. Whereas the nuclear magnetic
resonance experiments probe the local environment of one
particular nucleus and therefore a wavevector average of the dynamical
structure factor at very low energy transfers, the neutron scattering
experiments can, at least in principle, scan all frequencies and
wavevectors corresponding to the whole Brillouin zone.

The interpretation of the nuclear magnetic resonance experiments is quite
involved. It requires the knowledge of the so called hyperfine Hamiltonian
which describes the coupling of the electronic spin to the nuclear spin.
By now it established that the spin degrees of freedom in the CuO
superconductors can be described in a so called ``one component picture'',
which means that it is sufficient to consider one spin degree of freedom
per unit cell. If one assumes a band structure picture one can say more
precisely that only one band plays a role for the low energy physics of the
spins. The nature of the band, i.e. the percentage of copper or oxygen
admixture might vary, as one moves along the Fermi surface. But the
statement about one component is more general in that it does not
require a Fermi liquid state. The spin degree of freedom might equally
well be carried by the strongly correlated quasiparticles of a tJ - model.
The question is how do the spin of the quasiparticle couple to the
nuclear spins. This is discussed in the next section.

\section{
Transferred hyperfine interactions and band structure
}

In this section we discuss the derivation of a hyperfine Hamiltonian
similar to the so called Mila-Rice Hamiltonian \cite{Mila Rice}
which was originally introduced to understand the difference in
the anisotropies of the Cu nuclear magnetic
relaxation rates and the Cu Knight shift measurements in terms of
uncorrelated spins.
The importance of
a transferred hyperfine interaction in the case where the electronic spins
are strongly correlated was pointed out early on by Shastry \cite{Shastry PRL}.
In his paper he showed that within the theoretical framework of the so called
t-J model one can understand the difference between the magnitude of the
Cu and O relaxation rates if the oxygen nuclear spins are only coupled to the
electronic spins via a transferred hyperfine coupling.

The phenomenological hyperfine Hamiltonian which is compatible with
this assumption of a ``one component picture'' is:

\begin{equation}
H_{hf} =
\sum_{i\alpha\beta}
{^{63}{I}_{i\alpha\beta}} A_{\alpha\beta} {{S}_{i\beta}}
+ B \sum_{<ij>\alpha}
{^{63}{I}_{i\alpha}} {{S}_{j\alpha}}
+ \sum_{<ij>\alpha\beta}
{^{17}{I}_{i\alpha}} C_{\alpha\beta} {{S}_{j\beta}}
  \label{hyperfine hamiltonian}
\end{equation}

\noindent
where ${^{63}I_{i\alpha}}$ is the the $\alpha$th, $\alpha \in \{x,y, z\}$,
component of the copper nuclear spin at the site $i$,
$S_{i\alpha}$ is the $\alpha$th component of the electronic spin,
$A_{\alpha\beta}$ is the anisotropic copper hyperfine coupling, $B$ is the
transferred Cu hyperfine coupling and $C_{\alpha\beta}$ is the transferred
oxygen hyperfine coupling.

This is the so called Mila-Rice Hamiltonian. The unusual feature is
the large so called transferred hyperfine coupling $B$. Couplings to
other nuclei can be included in a similar way.
Whereas the Mila-Rice Hamiltonian was derived from a quantum-chemistry
calculation on a small cluster and therefore in a very ionic picture Shastry
assumed an itinerant electron system. It was argued that the success of the
Mila-Rice Hamiltonian indicates that the Cu spins are nearly localized.
Here I would like to point out that in an itinerant ``band structure picture''
one basically obtains the same results for the hyperfine Hamiltonian. These
results are mostly due to Takigawa \cite{Takigawa}.
In addition to the Mila Rice Hamiltonian one finds additional terms which
are of
similar size as the nearest neighbor transferred hyperfine interaction but we
will argue that these terms are negligible because the correlations are only
short ranged.

The starting point of our calculation is the canonical band structure for the
plane. We include in addition to copper $3d_{x^2-y^2}$, the oxygen $2p_x$ and
$2p_y$ orbitals and the copper $4s$ orbital. Here it is sufficient to include
the $4s$ orbital only in perturbation theory. The Hamiltonian has the form
\begin{eqnarray}
H_0 &=&
\sum_{k\alpha}
\{
\epsilon_d \dd{k\alpha} \du{k\alpha}
+  \epsilon_p \pk{x,k\alpha} \pu{x,k\alpha}
+  \epsilon_p \pk{y,k\alpha} \pu{y,k\alpha}
\} \nonumber\\
&+&
\sum_{k\alpha}
2 i t \dd{k\alpha}
(
\sin(k_x/2) \pu{x,k\alpha} + \sin(k_y/2) \pu{y,k\alpha}
)
+ h.c.
  \label{band Hamiltonian}
\end{eqnarray}

\noindent
where $\dd{k\alpha}$ ($\du{k\alpha}$) is the copper $3d_{x^2-y^2}$
electron creation (annihilation) operator, $\epsilon_d$ is the energy of
that orbital, $\pk{x,y,k\alpha}$ are the annihilation operators for the
oxygen ($p_x$, $p_y$) orbitals in the plane.
If the $4s$ orbital is partially filled $4s$ the
electron spin has a contact interaction with the nuclear spin at site $i$ which
is given by:

\begin{equation}
H_{hf} = B \;\; {^{63}\vec I_i} \;\cdot\;
\sd{i\alpha}{\vec \sigma}_{\alpha\beta} \su{i\beta}
  \label{contact term}
\end{equation}

\noindent
where $\sd{i\alpha}$ are the annihilation (creation) operators
for the $4s$ orbital at the Cu site $i$, $\vec \sigma$ is the vector of Pauli
matrices and ${^{63}\vec I_i}$ is the Cu nuclear spin. The coupling constant
$B$ depends on atomic physics and can be found for example in
\cite{Mila Rice}. The contact interaction is isotropic and therefore the
transferred
hyperfine coupling will be isotropic. If we believe that only one band of the
three bands of our model plays a role for the low energy physics we would
like to reduce hyperfine Hamiltonian to the part which is connected with
this one band. Therefore we project the hyperfine Hamiltonian,
Eq. \ref{contact term}, on to the conduction electron band:

\begin{equation}
\left(H_{hf}\right)_{proj}=\sum_{kk'}
<k\alpha| H_{hf} |k'\alpha'> \cd{k\alpha} \cu{k'\alpha'}
  \label{projection}
\end{equation}

\noindent
where $\cd{k\alpha}$ are the creation and annihilation operators
of the conduction electrons with momentum $\vec k$ and spin $\alpha$.

The projected hyperfine Hamiltonian has the form:

\begin{equation}
\left( H_{hf} \right)_{proj} = B\;
{^{63}{\vec I}}_0
\cdot
\sum_{ij} f^*_i f_j
\cd{i\alpha} {\vec \sigma}_{\alpha\beta} \cu{j\beta}
  \label{projected Hamiltonian}
\end{equation}

\noindent
where $f_i$ is the Fourier transform of $f_k = <4s|k>$, the overlap between
the $4s$ state and the conduction
electron bands at the site $i$.
It turns out that we only need to know this overlap. This is
basically what Mila and Rice did for their ionic model.
The parameters for a fit of tight binding model to the ``real band
structure'' were given by e.g.  Hybertsen and Schl\"uter \cite{Hybertsen}.
For the energy of the Cu $d$
orbital, the oxygen $p$ orbital and the Cu $4s$ orbital respectively they
found $\epsilon_d = -2.1 {\rm eV}$, $\epsilon_p = -3.3 {\rm eV}$ and
$\epsilon_{4s} = -8.1 {\rm eV}$. For the overlaps they give the following
numbers $t_{pd} = 1.3 {\rm eV}$, $t_{pp} = 1 {\rm eV}$ and
$t_{ps} = 3.8{\rm eV}$.
With this parameters we can calculate the hyperfine interactions (for
details consult e.g. \cite{Millis Monien}).
In addition to the terms which couple the conduction electron spin on one site
to the nuclear spin on a neighboring site
we find terms which describe a nuclear spin assisted spin
flip + hop term. The result of the calculation can be cast into the simple
form:

\begin{equation}
\left( H_{hf} \right)_{proj}
= B_{10,10} {^{63}\vec{I}_{00}} \cdot \vec S_{11}
+ B_{10,01} {^{63}\vec{I}_{00}} \cdot
\cd{10,\alpha}{\vec \sigma}_{\alpha\beta}\cu{01,\beta} + ...
\end{equation}

\noindent
where the coupling constants are of the order
$B_{10,10}\sim 48 {\rm kOe}/\mu_B$, $B_{10,01} = B_{10,10}$.
The  coupling constant $B_{10,10}$ is the transferred hyperfine coupling for
the Cu nuclear spin as disussed by Mila and Rice, the spin flip hop term
$B_{10,01}$ is not present in the ionic picture. The next term in the
expansion is about a factor of 6 smaller then the first term and can be
neglected savely.
Although the nuclear spin assisted spin flip hop term is of the
same magnitude as the transferred term we can neglect it in our analysis of
the nuclear magnetic resonance experiments because the dynamical structure
factor does not have large weight at this particular area of momentum space.
Therefore we are left with the standard hyperfine Hamiltonian for the CuO
plane.

If we do not allow for any other relaxation channel which might for example
be produced by a coupling of the nuclear spin to the orbital motion of the
electron then we know exactly which moments of the dynamical structure factor
contribute to the nuclear magnetic relaxation.

\section{
How important is orbital relaxation?
}

In previous theories of the nuclear magnetic relaxation in the high temperature
superconductors it was assumed the only way the nuclear spin couples to the
electrons is via the electronic spin.
The unusual behavior of  the anisotropy
of the Cu nuclear magnetic relaxation rate
as a function of temperature \cite{Takigawa Cu anisotropy},
\cite{Slichter Cu anisotropy}
raises the question if there is
another relaxation channel present.
In a recent paper Millis and Monien
\cite{Millis Monien orbital relaxation} discuss the possibility of another
relaxation channel for the Cu nuclear spin, namely the coupling of the Cu
nuclear spin to the orbital motion of the electron which was already
mentioned by Warren and Walstedt in an early paper. The coupling of the
nuclear spin to the orbital momentum is give by:

\begin{equation}
H_{\rm Orb} = {^{63}\gamma_n}\gamma_e\hbar^2
\frac{{\vec I}\cdot{\vec L}}{r^3}
\label{Horb}
\end{equation}

\noindent
where $\vec I$ is the copper
nuclear spin, $\vec L$ the electron angular momentum
and ${^{63}\gamma}_n$ and $\gamma_e$ the gyromagnetic ratio of the copper
nucleus
and the electron respectively.
In order to estimate the size of this effect we use again the simplest
tight binding picture for the CuO band of a single plane with the same
parameter as used for the calculation of the hyperfine Hamiltonian.
For the calculation of the relaxation rates we need to know the
matrix elements of the orbital coupling, Eq. (\ref{Horb}), with the
Wannier functions. The rates can then be calculated simply by using
Fermis ``golden rule''. At a temperature of about 100K we find for the
nuclear magnetic relaxation rate produced by {\em orbital currents} for a field
applied parallel and perpendicular to the CuO plane:

\begin{eqnarray}
W_\parallel
&=& 3.8 \times C_{x^2-y^2} \left(4 C_{xy} + C_{xz}\right) \ {\rm ms^{-1}}
\cr
W_\perp
&=& 3.8 \times C_{x^2-y^2} C_{xz} \ {\rm ms^{-1}}
  \label{Eq: Absolute values}
\end{eqnarray}

The coefficients $C_{x^2-y^2}$, $C_{xy}$ and $C_{xz}$ denote the percentage
of admixture of the Cu $3d_{x^2-y^2}$, $3d_{xy}$ and $3d_{xz}$ orbital
respectively.
The result of the analysis is that the orbital relaxation
of the nuclear spins can be appreciable because of the orbital degeneracy
but that the absolute value depends strongly on the exact amount of admixture
of the Cu $3d_{xz}$ and $3d_{xy}$ orbitals to the Wannier functions at the
Fermi surface which are not known theoretically
to an accuracy, ($\sim 5\%$),
needed to determine if the orbital relaxation may play a role or not.
Nevertheless it is important to keep this additional relaxation channel
in mind if one discusses the limitations on hyperfine Hamiltonian and
dynamical structure factor obtained from experiments.

\section{
Reexamination of the MMP model
}
Up to now we have only discussed the coupling of the electronic
spin and orbital degree of freedom to the nuclear spin. Now we would
like to turn our attention to the quantity which is of most interest
from the theoretical point of view, the dynamical structure factor.

Recently we \cite{Millis Monien} reexamined the results of a
previous phenomenological analysis of the nuclear magnetic relaxation
experiments in the \YBCO{6+\delta} materials \cite{MMP}.
In the
original MMP paper we made the assumption that the only temperature
dependent quantity is the antiferromagnetic correlation length
and the energy scale is described by a dynamical critical exponent
and used the simplest form for a peaked structure factor describing the
magnetic correlations, a Lorentzian.
In the extended analysis we dropped the assumption
that the correlation length is the only quantity which is temperature dependent
and allowed the strength of the antiferromagnetic peak also to vary
with temperature. Also we considered two models for the $q$ dependence of the
structure factor - a Lorentzian and a Gaussian. The parameters describing the
correlation functions are the antiferromagnetic correlation length $\xi$ and
the strength of the antiferromagnetic peak which we denote with $\beta$. The
two models for the dynamical structure factor in the limit
frequency $\omega$ going to zero are:
\begin{eqnarray}
{\chi^{''}}_{Lor}(q, \omega) &=& \frac{\pi\chi_0\hbar\omega}{\Gamma}
\left[ 1 + \beta \frac{\xi^4}{(1+\xi^2 (q - Q)^2)^2} \right] \\
{\chi^{''}}_{Gauss}(q, \omega) &=& \frac{\pi\chi_0\hbar\omega}{\Gamma}
\left[ 1 + \beta \xi^4 e^{ - \xi^2 (q - Q)^2 } \right]
  \label{chi models}
\end{eqnarray}
where $\xi$ is the antiferromagnetic correlation length, $\beta$ is the
strength of the antiferromagnetic peak at the zone corner, $Q=(\pi/a,\pi/a)$,
$\Gamma$ is
the ``bare'' energy scale of the spin fluctuations and $\chi_0$ is the
static susceptibility.

Within the ``one component
picture'' the Knight shift for a field applied in the direction $\alpha$,
$K_\alpha$, and the relaxation rates, $W_\alpha$ of the various nuclei,
denoted by $a = 63, 17, 89$ for Cu, O and Y respectively,
for a field applied in the $\alpha$ direction are given by:

\begin{eqnarray}
{^a K}_\alpha &=& \lim_{q \rightarrow 0} {^a F}_\alpha(q)\chi'(q,\omega=0)
\nonumber \\
{^a W}_\alpha &=& \frac{1}{4\mu_B^2} \lim_{\omega \rightarrow 0}
\frac{k_B T}{\omega}
\sum_{q,{\overline \alpha}}
\left[{^a F}_{{\overline \alpha}}\right]^2 \chi''(q,\omega)
  \label{basic NMR}
\end{eqnarray}

\noindent
where ${\overline \alpha}$ is the direction perpendicular to $\alpha$. The
form factors ${^a F}_\alpha(q)$ have the dimension of energy and are
basically the Fourier transform of the spin hyperfine Hamiltonian and
can be found in the literature, e.g. \cite{Millis Monien}.
To
obtain the limits on the correlation lengths we considered two limits in one
case the correlation length is temperature dependent and the strength is not.
In the second case the strength of the peak is temperature dependent and the
correlation length is temperature independent.

We state the results for the correlation lengths in
Table (\ref{tab: correlation lengths}).
\begin{table}[h]
  \leavevmode
  \begin{center}
    \begin{tabular}{|c|c|c|c|c|}
        \cline{2-5}
          \multicolumn{1}{c} {\ }
        & \multicolumn{2}{|c|}{\YBCO{6.63}}
        & \multicolumn{2}{c|}{\YBCO{7}} \\ \hline
      T[K] & $\xi_{\rm Lor}$ & $\xi_{\rm Gauss}$
           & $\xi_{\rm Lor}$ & $\xi_{\rm Gauss}$ \\ \hline
      100  & 1.5 - 2.4 & 1.0 - 1.5 & 2.5 - 4.5 & 1.5 - 2.5 \\ \hline
      300  & 1.3 - 1.7 & 0.7 - 1.1 & 1.5 - 2.5 & 1.0 - 1.5 \\ \hline
    \end{tabular}
  \caption{correlation lengths from the NMR analysis}
  \label{tab: correlation lengths}
  \end{center}
\end{table}

\noindent
 From these results we can conclude:
\begin{itemize}
\item The results for the correlation length are not compatible with the
  neutron scattering results by Rossat - Mignod \cite{Rossat - Mignod}.
\item The neutron scattering experiments favor temperature independent
  correlation lengths.
\item The experimental results by G. Aeppli and coworkers
\cite{Aeppli LaSrCuO} seem to be more
  compatible with the correlation lengths required by NMR experiments.
\end{itemize}
The detailed analysis shows that the copper and oxygen relaxation time
experiments are compatible with both assumptions - a temperature
dependent correlation length or a temperature dependent strength of the
antiferromagnetic peak - but that the crucial quantity which could determine
which case it is is the ratio of the yttrium to the oxygen relaxation rate
which was not measured in the same sample to a high enough accuracy
up to now. For details of the analysis we refer the reader to our paper
\cite{Millis Monien}.
The original MMP analysis was intended to describe
the spin correlations on a very low energy scale. A comparison of the
nuclear magnetic experiments with the neutron scattering experiments requires
an extrapolation to larger energies. For a thorough discussion we refer
the reader to \cite{Millis Monien}

\section{
Dynamical structure factor for in the marginal Fermi liquid picture
}
The remaining physics question, as stated in the introduction, is off course
where does one get the low energy physics scale from. The easiest way to
obtain a low energy scale from a ``high energy'' model is to be close
to a critical point in this class of models belong the ``nested Fermi liquid'',
Hubbard RPA and dynamical phase separation models. Very close to the phase
transition dynamical scaling should hold in that case the low energy scale
is determined by the correlation length via $\omega \sim \Gamma^{-z}$ where
$\Gamma$ is the energy scale determined by the ``high energy physics'' and
$z$ is the dynamical scaling exponent. In this section we will circumnavigate
the difficult question what is determining the low energy physics and answer
the following related
question: Assuming that the only effect of all the many body
complications is to give rise to a single particle scattering rate which
is proportional to the energy or temperature, whatever is less (which
is the so called marginal Fermi liquid hypothesis) what can
we know about the spin response.
The original marginal Fermi liquid picture, \cite{MFL}, assumed that the
selfenergy is nearly wavevector independent
\begin{equation}
{\rm Im}\Sigma \sim {\rm max}( \omega, T).
  \label{MFL selfenergy}
\end{equation}
In the original MFL paper it was argued that the resulting response
functions are a universal function of $\omega/T$ where $\omega$ is the
frequency of the external field and $T$ is the temperature and also weakly
$q$ dependent. The assumed factorizable form of the dynamical structure
factor is not consistent with a different temperature dependence of the
Cu and O relaxation rate.
Neutron scattering experiments by Aeppli et al. \cite{Aeppli LaSrCuO}
on the \LaSrCuO for $x\approx 0.15$ showed that the dynamical structure
factor is strongly $q$ dependent even for the strongly doped material.
Littlewood et al., \cite{Littlewood et al.} have
proposed an extension of the original MFL picture.
It is clear that in two dimensions the dynamical structure factor even for the
noninteracting Fermi gas shows a strong $q$ dependence at low frequencies.
It is interesting to examine the predictions of the simple band structure
of the CuO plane. Using a tight binding band structure one obtains an
incommensurate peak structure with four posts which are a little bit
shifted away from the zone corner, $Q=(\pi/a,\pi/a)$.
What we found is that the position and magnitude seems to
be compatible with a simple tight binding band structure picture for the CuO
plane. The position and magnitude of the peaks will strongly depend on
the doping. What is even more interesting is that if we include the
marginal Fermi liquid form for the self energy we also can understand
the temperature and energy dependence of the peak structure.
In the regime
which is accessible by the neutron scattering the main effect of the
marginal Fermi liquid form of the quasiparticle propagator is the life
time broadening which has the energy scale of the temperature.

\begin{figure}[h]
    \SmallProceedingsFigure{sq.ps}
    \caption{The dynamical structure factor $S(q,\omega)$ for various
    energy transfers, $\omega=$ 3, 6, 12 and 15 meV as a function of
    the momentum transfer in the $q_x$ direction. $q_y$ is fixed to $\pi/a$.
    The bottom curve corresponds to the lowest energy transfer the top
    to the highest.}
    \label{fig:1}
\end{figure}

The magnetic
properties of the heavily doped regime of \LaSrCuO can therefore be understood
in terms of a simple band structure picture of marginal Fermi liquid
quasiparticles.

\section{Conclusions}

By now it has become quite clear that temperature dependent antiferromagnetic
correlations play an
important role in the understanding of the magnetic properties of the
CuO high temperature superconductors even in the strongly doped regime.
The early interpretation of Millis, Monien and Pines \cite{MMP}
assumed a temperature dependent correlation length. The reexamination
of the present nuclear magnetic resonance data by \cite{Millis Monien}
demonstrates that one can not deduce from the nuclear magnetic relaxation
measurements of the copper and oxygen rates alone that the correlation length
is temperature dependent. We have proposed a simple model in which the
temperature and frequency dependence of the dynamical structure factor
in \LaSrCuO, $x\approx 0.15$, can be understood in a simple tight binding
band structure picture including lifetime effects.

It is a pleasure to thank my numerous collaborators. Many have worked
in this interesting area of solid state physics - I apologize to those
whose important work I did omit in this brief review. The research for this
project was supported in part by an IBM fellowship
and by the NSF Grant NSF-PHY-04035.

\end{document}